\PassOptionsToPackage{usenames}{color}
\pdfoutput=1 
\documentclass[10pt, a4paper]{article}
\usepackage{lrec}

\makeatletter
\renewcommand{\p@section}{\arabic{section}\expandafter\@gobble}
\renewcommand{\p@subsection}{\thesection\arabic{subsection}\expandafter\@gobble}
\renewcommand{\p@subsubsection}{\thesubsection\arabic{subsubsection}\expandafter\@gobble}
\makeatother

\usepackage{nert}

\usepackage{graphicx}
\usepackage{url} 
\usepackage{times}

\usepackage{tabularx}
\usepackage{soul}
\usepackage{microtype}
\usepackage{xcolor}
\usepackage{linguex}
\usepackage{amsmath}
\usepackage{epstopdf}
\usepackage{microtype}
\usepackage{latexsym}
\usepackage{listings}
\usepackage{multicol}

\usepackage{xstring}
\usepackage{pinyin}
\usepackage[T1]{fontenc}
\usepackage{CJKutf8}
\usepackage{enumitem}

\newcommand\pos[1]{\textsc{#1}}

\title{\textbf{A Corpus of Adpositional Supersenses for Mandarin Chinese}}

\name{Siyao Peng, Yang Liu, Yilun Zhu, Austin Blodgett, Yushi Zhao, Nathan Schneider}

\address{Department of Linguistics, Georgetown University \\
         Washington, DC, USA \\
         \{sp1184, yl879, yz565, ajb341, yz521, nathan.schneider\}@georgetown.edu\\}

\definecolor{myorange}{RGB}{255, 127, 14}
\definecolor{myblue}{RGB}{31, 119, 180}

\abstract{
Adpositions are frequent markers of semantic relations, but they are highly ambiguous and vary significantly from language to language. 
Moreover, there is a dearth of annotated corpora for investigating the cross-linguistic variation of adposition semantics, or for building multilingual disambiguation systems. 
This paper presents a corpus in which all adpositions have been semantically annotated in Mandarin Chinese; to the best of our knowledge, this is the first Chinese corpus to be broadly annotated with adposition semantics.
Our approach adapts a framework that defined a general set of \emph{supersenses} according to ostensibly language-independent semantic criteria, though its development focused primarily on English prepositions  \citep{schneider-etal-2018-comprehensive}.
We find that the supersense categories are well-suited to Chinese adpositions despite syntactic differences from English.
On a Mandarin translation of \textit{The Little Prince}, 
we achieve high inter-annotator agreement and analyze semantic correspondences of adposition tokens in bitext. 
 \\ \newline 
}

\begin{document}

\maketitleabstract

\section{Introduction}

Adpositions (i.e.~prepositions and postpositions) include some of the most frequent words in languages like Chinese and English, 
and help convey a myriad of semantic relations 
of space, time, causality, possession, and other domains of meaning.
They are also a persistent thorn in the side of second language learners owing to their extreme idiosyncrasy \citep{chodorow-07,lorincz2012difficulties}.
For instance, the English word \emph{in} has no exact parallel in another language; rather, for purposes of translation, its many different usages cluster differently depending on the second language.
Semantically annotated corpora of adpositions in multiple languages, including parallel data, would facilitate broader empirical study of adposition variation than is possible today, 
and could also contribute to NLP applications such as machine translation \citep{li-05,agirre-09,shilon-12,weller-14,weller-15,hashemi-14,popovic-17} and grammatical error correction \citep{chodorow-07,tetreault-08,de_felice-08,hermet-09,huang-16,graen-17}.
\\[5pt]
This paper describes the first corpus with broad-coverage annotation of adpositions in Chinese. 
For this corpus we have adapted \citeposs{schneider-etal-2018-comprehensive} Semantic Network of Adposition and Case Supersenses annotation scheme (SNACS; see \cref{sec:snacs}) to Chinese.\footnote{\citet{zhu2019adpositional} previewed our approach.} Though other languages were taken into consideration in designing SNACS, no serious annotation effort has been undertaken to confirm empirically that it generalizes to other languages.
After developing new guidelines for syntactic phenomena in Chinese (\cref{subsec:adposition_criteria}), we apply the SNACS supersenses to a translation of \emph{The Little Prince}\footnote{Originally \emph{Le Petit Prince} by Antoine de St.~Exup\'{e}ry, published in 1943 and subsequently translated into numerous languages.} (\textit{\Xiao3 \Wang2 \Zi3}), 
finding the supersenses to be robust and achieving high inter-annotator agreement (\cref{sec:corpus-annotation}).
We analyze the distribution of adpositions and supersenses in the corpus, and compare to adposition behavior in a separate English corpus (see \cref{sec:corpus-analysis}).
We also examine the predictions of a part-of-speech tagger in relation to our criteria for annotation targets (\cref{sec:adpositionidentification}).
The annotated corpus and the Chinese guidelines for SNACS will be made freely available online.\footnote{\url{https://github.com/nert-nlp/Chinese-SNACS/}}

\section{Related Work} \label{related-work}


To date, most wide-coverage semantic annotation of prepositions has been dictionary-based, taking a word sense disambiguation perspective  \citep{litkowski2005preposition, litkowski-hargraves-2007-semeval, litkowski-2014-pattern}.
\Citet{schneider-15-hierarchy} proposed a supersense-based (unlexicalized) semantic annotation scheme which would be applied to all tokens of prepositions in English text. We adopt a revised version of the approach, known as SNACS (see \cref{sec:snacs}). Previous SNACS annotation efforts have been mostly focused on English---particularly STREUSLE \citep{schneider2016corpus,schneider-etal-2018-comprehensive}, the semantically annotated corpus of reviews from the English Web Treebank \citep[EWT;][]{bies2012english}.
We present the first adaptation of SNACS for Chinese by annotating an entire Chinese translation of \textit{The Little Prince}.


\subsection{Chinese Adpositions and Roles}

In the computational literature for Chinese, apart from some focused studies (e.g., \citet{yang-98} on logical-semantic representation of temporal adpositions), there has been little work addressing adpositions specifically. 
Most previous semantic projects for Mandarin Chinese focused on content words and did not directly annotate the semantic relations signaled by functions words such as prepositions \citep{xue-etal-2014-interlingua, hao2007description, you2005building, li-etal-2016-annotating}.
For example, in Chinese PropBank, \citet{xue2008labeling} argued that 
the head word and its part of speech are clearly informative for labeling the semantic role of a phrase, but the preposition is not always the most informative element. \citet{li-03} annotated the Tsinghua Corpus \citep{zhang1999language} from \textit{People’s Daily} where the content words were selected as the headwords, i.e., the object is the
headword of the prepositional phrase. In these prepositional phrases, the nominal headwords were labeled with one of the 59 semantic relations (e.g.~\textit{Location, LocationIni, Kernel word}) whereas the prepositions and postpositions were respectively labeled with syntactic relations \textit{Preposition} and \textit{LocationPreposition}.\footnote{Though named \textit{LocationPreposition} in \citet{li-03}, these adpositions actually occur postnominally, equivalent to localizers in this paper.}
Similarly, in Semantic Dependency Relations (SDR, \citealt{che-12,che-16}), prepositions and localizers were labeled as semantic markers  \textit{mPrep} and \textit{mRange}, whereas semantic roles, e.g., \textit{Location, Patient}, are assigned to the governed
nominal phrases.
\\[5pt]
\Citet{sun-jurafsky-2004-shallow} compared PropBank parsing performance on Chinese and English, and showed that four Chinese prepositions (\textit{\zai4},  \textit{\yu2}, \textit{\bi3}, and \textit{\dui4}) are among the top 20 lexicalized syntactic head words in Chinese PropBank, bridging the connections between verbs and their arguments.
The high frequency of prepositions as head words in PropBank reflects their importance in context. However, very few annotation scheme attempted to directly label the semantics of these adposition words.
\\[5pt]
\Citet{CKIP-report-1993} is the most relevant adposition annotation effort, categorizing Chinese prepositions into 66 types of senses
grouped by lexical items. 
However, these lexicalized semantic categories are constrained to a given language and a closed set of adpositions. 
For semantic labeling of Chinese adpositions in a multilingual context, we turn to the SNACS framework, described below.




\subsection{SNACS: Adposition Supersenses}\label{sec:snacs}

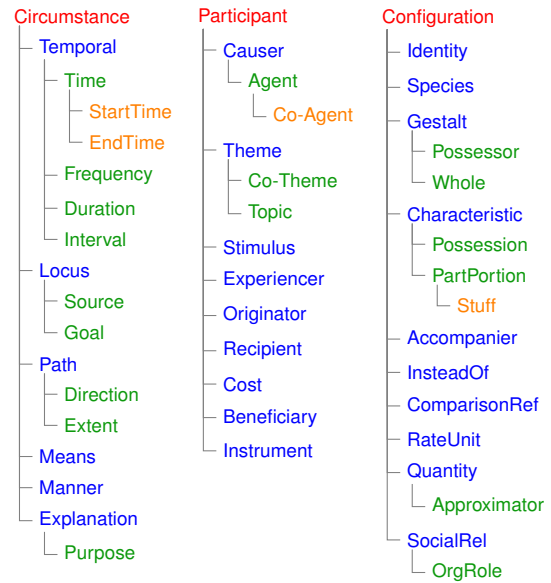
\begin{figure}[ht]
    \centering
    \newcommand{\hierA}[3]{\textcolor{red}{#1}}
\newcommand{\hierB}[3]{\textcolor{blue}{#1}}
\newcommand{\hierC}[3]{\textcolor{mdgreen}{#1}}
\newcommand{\hierD}[3]{\textcolor{orange}{#1}}
\newenvironment{ggroup}{{}}{{}}
\begin{minipage}{.80\columnwidth}
\vspace{.2cm}
\begin{multicols}{3}
\begin{ggroup}
  \sffamily\smaller\color{gray}
\begin{forest}
  for tree={%
    folder,
    grow'=0,
    fit=band,
    inner ysep=.75,
  }
  [{\hierA{Circumstance}{76, 63}{77}}
    [{\hierB{Temporal}{0}{0}}
      [{\hierC{Time}{360, 329}{371}}
        [{\hierD{StartTime}{28, 28}{28}}]
        [{\hierD{EndTime}{30, 31}{31}}]
      ]
      [{\hierC{Frequency}{9, 7}{9}}]
      [{\hierC{Duration}{90, 87}{91}}]
      [{\hierC{Interval}{4, 35}{35}}]
    ]
    [{\hierB{Locus}{636, 780}{846}}
      [{\hierC{Source}{77, 189}{189}}]
      [{\hierC{Goal}{234, 378}{419}}]
    ]
    [{\hierB{Path}{26, 44}{49}}
      [{\hierC{Direction}{120, 160}{161}}]
      [{\hierC{Extent}{42, 38}{42}}]
    ]
    [{\hierB{Means}{17, 16}{17}}]
    [{\hierB{Manner}{134, 48}{140}}]
    [{\hierB{Explanation}{121, 108}{123}}
      [{\hierC{Purpose}{301, 396}{401}}]
    ]
  ]
\end{forest}
\columnbreak

\begin{forest}
  for tree={%
    folder,
    grow'=0,
    fit=band,
    inner ysep=.75,
  }
  [{\hierA{Participant}{0}{0}}
    [{\hierB{Causer}{9, 10}{15}}
      [{\hierC{Agent}{158, 37}{170}}
        [{\hierD{Co-Agent}{35, 65}{65}}]
      ]
    ]
    [{\hierB{Theme}{224, 177}{238}}
      [{\hierC{Co-Theme}{14, 7}{14}}]
      [{\hierC{Topic}{213, 289}{296}}]
    ]
    [{\hierB{Stimulus}{123, 0}{123}}]
    [{\hierB{Experiencer}{107, 0}{107}}]
    [{\hierB{Originator}{134, 0}{134}}]
    [{\hierB{Recipient}{122, 0}{122}}]
    [{\hierB{Cost}{48, 30}{48}}]
    [{\hierB{Beneficiary}{93, 76}{110}}]
    [{\hierB{Instrument}{23, 19}{30}}]
  ]
\end{forest}
\columnbreak

\begin{forest}
  for tree={%
    folder,
    grow'=0,
    fit=band,
    inner ysep=.75,
  }
  [{\hierA{Configuration}{0}{0}}
    [{\hierB{Identity}{64, 77}{85}}]
    [{\hierB{Species}{39, 39}{39}}]
    [{\hierB{Gestalt}{165, 699}{709}}
      [{\hierC{Possessor}{381, 489}{492}}]
      [{\hierC{Whole}{142, 173}{250}}]
    ]
    [{\hierB{Characteristic}{133, 66}{140}}
      [{\hierC{Possession}{21, 2}{21}}]
      [{\hierC{PartPortion}{56, 36}{57}}
        [{\hierD{Stuff}{17, 25}{25}}]
      ]
    ]
    [{\hierB{Accompanier}{28, 47}{49}}]
    [{\hierB{InsteadOf}{10, 9}{10}}]
    [{\hierB{ComparisonRef}{176, 184}{215}}]
    [{\hierB{RateUnit}{5, 5}{5}}]
    [{\hierB{Quantity}{191, 84}{191}}
      [{\hierC{Approximator}{76, 73}{76}}]
    ]
    [{\hierB{SocialRel}{240, 0}{240}}
      [{\hierC{OrgRole}{103, 0}{103}}]
    ]
  ]
\end{forest}
\end{ggroup}
\end{multicols}
\end{minipage}
    \caption{SNACS hierarchy of 50 supersenses.}
    \label{fig:supersenses}
\end{figure}

\citet{schneider-etal-2018-comprehensive} proposed the Semantic Network of Adposition and Case Supersenses (SNACS), a hierarchical inventory of 50 semantic labels, i.e., supersenses, that characterize the use of adpositions, as shown in \Cref{fig:supersenses}.
Since the meaning of adpositions is highly affected by the context, SNACS can help distinguish different usages of adpositions. For instance, \cref{single-label} presents an example of the supersense \psst{Topic} for the adposition \textit{about} which 
emphasizes the subject matter of urbanization that the speaker discussed. In \cref{single-label-amb}, however, the same preposition \textit{about} takes a measurement in the context, expressing an approximation.

\ex. I gave a presentation \textbf{about:\psst{Topic}} urbanization.\footnote{Throughout this paper, adposition tokens under discussion are bolded and labeled.} \label{single-label}

\ex. We have \textbf{about:\psst{Approximator}} 3 eggs left. \label{single-label-amb}

Though assigning a single label to each adposition can help capture its lexical contribution to the sentence meaning as well as disambiguate its uses in different scenarios, the canonical lexical semantics of adpositions are often stretched to fit the needs of the scene in actual language use. 

\ex. I care \textbf{about:\rf{Stimulus}{Topic}} you. \label{eg:stimulustopic}

For instance, \cref{eg:stimulustopic} blends the domains of emotion (principally reflected in \emph{care}, which licenses a \psst{Stimulus}), and cognition (principally reflected in \emph{about}, which often marks non-emotional \psst{Topic}s). Thus, SNACS incorporates the \emph{construal analysis} \citep{hwang-etal-2017-double} wherein the lexical semantic contribution of an adposition (its \textbf{function}) is distinguished and may diverge from the underlying relation in the surrounding context (its \textbf{scene role}). Construal is notated by \rf{SceneRole}{Function}, as 
 \rf{Stimulus}{Topic} in \cref{eg:stimulustopic}.\footnote{The supersense labels in \emph{congruent} construals, such as \psst{Topic} and \psst{Approximator} in \Cref{single-label} and 
 \Cref{single-label-amb}, are both function and scene role by definition.}
\\[5pt]
Another motivation for incorporating the construal analysis, as pointed out by \citet{hwang-etal-2017-double}, is its capability to adapt the English-centric supersense labels to other languages, which is the main contribution of this paper. The construal analysis can give us insights into the similarities and differences of function and scene roles of adpositions across languages. 

\section{Adposition Criteria in Mandarin Chinese}\label{subsec:adposition_criteria}

Our first challenge is to determine which tokens 
qualify as adpositions in Mandarin Chinese and merit supersense annotations. 
The English SNACS guidelines (we use version~2.3) broadly define the set of SNACS annotation targets to include canonical prepositions (taking an noun phrase (NP) complement) and their subordinating (clausal complement) uses. Possessives, intransitive particles, and certain uses of the infinitive marker \emph{to} are also included \citep{schneider2018guideline}.
\\[5pt]
In Chinese, the difficulty lies in two areas, which we discuss below. Firstly, prepositional words are widely attested. However, since no overt derivational morphology occurs on these prepositional tokens (previously referred to as coverbs), we need to filter non-prepositional uses of these words. Secondly,  post-nominal particles, i.e., localizers, though not always considered adpositions in Chinese, deliver rich semantic information.

\paragraph{Coverbs}
Tokens that are considered generic prepositions can co-occur with the main predicate of the clause and introduce an NP argument to the clause \citep{li1974co} as in \cref{zho:shangtopic}. These tokens are referred to as coverbs. 
In some cases, coverbs can also occur as the main predicate. For example, the coverb \textit{\zai4} heads the predicate phrase in \cref{zho:pred}.

\exg. \ta1 \textbf{\zai4:\psst{Locus}} \xue2\shu4 \textbf{\shang4:\rf{Topic}{Locus}}  \you3\suo3\zuo4\wei2.\\
\textsc{3sg} \textsc{p}:at academia \textsc{lc}:on-top-of successful \\
`He succeeded in academia.’ \label{zho:shangtopic}

\exg. \ni3 \yao4 de \yang2 \jiu4 \textbf{\zai4} \li3\mian4.\\
\textsc{2sg} want \textsc{de} sheep \textsc{res} at inside\\
`The sheep you wanted is in the box.' \label{zho:pred}
(\texttt{zh\_lpp\_1943.92}) 

In this project, we only annotate coverbs when they do not function as the main predicate in the sentence, echoing the view that coverbs modify events introduced by the predicates, rather than establishing multiple events in a clause \citep{hui2012order}. Therefore, lexical items such as \textit{\zai4} are annotated when functioning as a modifier as in \cref{zho:shangtopic}, but not when as the main predicate as in \cref{zho:pred}.

\paragraph{Localizers}

Localizers are words that follow a noun phrase to refine its semantic relation. For example, \textit{\shang4} in \cref{zho:shangtopic} denotes a contextual meaning, `in a particular area,' whereas the co-occurring coverb \textit{\zai4} only conveys a generic location.
It is unclear whether localizers are syntactically postpositions, but we annotate all localizers because of their semantic significance.
Though coverbs frequently co-occur with localizers and the combination of coverbs and localizers is very productive, there is no strong evidence to suggest that they are circumpositions. As a result, we treat them as separate targets for SNACS annotation: for example, \textit{\zai4} and \textit{\shang4} receive \psst{Locus} and \rf{Topic}{Locus} respectively in \cref{zho:shangtopic}.
\\[5pt]
Setting aside the syntactic controversies of coverbs and localizers in Mandarin Chinese, we regard both of them as adpositions that merit supersense annotations.
As in 
\cref{zho:shangtopic}, 
both the coverb \textit{\zai4} and the localizer \textit{\shang4} surround an NP argument \textit{\xue2\shu4} (`academia') and they as a whole modify the main predicate \textit{\you3\suo3\zuo4\wei2} (`successful').
In this paper, we take the stance that coverbs co-occur with the main predicate and precede an NP, whereas localizers follow a noun phrase and add semantic information to the clause.

\section{Corpus Annotation} \label{sec:corpus-annotation}

We chose to annotate the novella \textit{The Little Prince} because it has been translated into hundreds of languages and dialects, which enables comparisons of linguistic phenomena across languages on bitexts.
This is the first Chinese corpus to undergo SNACS annotation.
Ongoing adpositional supersense projects on \textit{The Little Prince} include English, German, French, and Korean. In addition, \textit{The Little Prince} has received large attention from other semantic frameworks and corpora, including the English \citep{banarescu2013abstract} and Chinese  \citep{li-etal-2016-annotating} AMR corpora.

\subsection{Preprocessing}

We use the same Chinese translation of \textit{The Little Prince} as the Chinese AMR corpus \citep{li-etal-2016-annotating}, which is also sentence-aligned with the English AMR corpus \citep{banarescu2013abstract}. These bitext annotations in multiple languages and annotation semantic frameworks can facilitate cross-framework comparisons. 
\\[5pt]
Prior to supersense annotation, we conducted the following preprocessing steps in order to identify the adposition targets that merit supersense annotation.

\paragraph{Tokenization}  After automatic tokenization using Jieba,\footnote{\url{https://github.com/fxsjy/jieba}} we conducted manual corrections 
to ensure that all potential adpositions occur as separate tokens, closely following the Chinese Penn Treebank segmentation guidelines
\citep{xia2000segmentation}. The final corpus includes all 27 chapters of \textit{The Little Prince}, with a total of 20k tokens. 


\paragraph{Adposition Targets}
All annotators jointly identified adposition targets according to the criteria discussed in \cref{subsec:adposition_criteria}. 
Manual identification of adpositions was necessary as an automatic POS tagger was found unsuitable for our criteria (\cref{sec:adpositionidentification}).

\paragraph{Data Format} Though parsing is not essential to this annotation project, we ran the StanfordNLP \citep{qi2018universal} dependency parser to obtain POS tags and dependency trees. 
These are stored alongside supersense annotations in the \mbox{\textit{CoNLL-U-Lex}} format \citep[modeled after the STREUSLE corpus;][]{schneider-15,schneider-etal-2018-comprehensive}. \mbox{CoNLL-U-Lex} extends the \mbox{CoNLL-U} format used by the Universal Dependencies \citep[UD;][]{nivre-16} project to add additional columns for lexical semantic annotations.\footnote{\url{https://github.com/nert-nlp/streusle/blob/master/CONLLULEX.md}}


\subsection{Reliability of Annotation}

The corpus is jointly annotated by three native Mandarin Chinese speakers, all of whom have received advanced training in theoretical and computational linguistics. 
Supersense labeling was performed cooperatively by 3~annotators for 25\% (235/933) of the adposition targets, and for the remainder, independently by the 3~annotators, followed by cooperative adjudication.
Annotation was conducted in two phases, and therefore we present two inter-annotator agreement studies to demonstrate the reproducibility of SNACS and the reliability of the adapted scheme for Chinese. 
\\[5pt]
\begin{table}[ht]
    \centering\small
    \begin{tabular}{lccc}
    \multicolumn{4}{c}{\textsc{IAA Samples}} \\ 
    \textbf{Phase} & \textbf{Time} & \textbf{Chapters} & \textbf{\# Adpositions} \\ \hline
    \multicolumn{1}{c}{Phase 1} & July 2018 & 15--20 & 111 \\
    \multicolumn{1}{c}{Phase 2} & Sept 2019 & 26--27 & 124 \\[2pt] \hline\hline
    \multicolumn{4}{c}{\textsc{Raw Agreement}} \\ 
    \textbf{Phase} & \textbf{Scene} & \textbf{Function} & \textbf{Construal} \\ \hline
    Phase 1 & .92 & .95 & .90 \\ 
    Phase 2 & .93 & .90 & .89 \\[2pt] \hline\hline
    \multicolumn{4}{c}{\textsc{Kappa}} \\ 
    \textbf{Phase} & \textbf{Scene} & \textbf{Function} & \textbf{Construal} \\ \hline
    Phase 1 & .90 & .93 & .88 \\ 
    Phase 2 & .92 & .88 & .88 \\ 
    \end{tabular}
    \caption{Inter-annotator agreement (IAA) results on two samples from different phases of the project. }
    \label{tab:iaa-results}
\end{table}
\Cref{tab:iaa-results} shows raw agreement and Cohen's kappa across three annotators computed by averaging three pairwise comparisons. Agreement levels on scene role, function, and full construal are high for both phases, attesting to the validity of the annotation framework in Chinese. However, there is a slight decrease from Phase 1 to Phase 2, possibly due to the seven newly attested adpositions in Phase 2 and the 1-year interval between the two annotation phases. 

\section{Corpus Analysis} \label{sec:corpus-analysis}

\begin{table}[t]
    \centering\small
    \begin{tabular}{l@{}r@{\hspace{5pt}}r}
         & \textbf{Toks.} & \textbf{Types} \\
         \midrule
         Chapters & 27 & NA \\
         Sentences & 1,597 & NA \\
         Tokens & 20,287 & NA \\ \midrule
         Adpositions & 933 & 70 \\
         \hspace{1em} Prepositions & 667 & 42 \\
         \hspace{1em} Postpositions & 266 & 28 \\
        \midrule
        Supersenses & 933 & 29 \\ 
         \hspace{1em} Scene roles & 933 & 28 \\
         \hspace{1em} Functions & 933 & 26 \\ 
         \midrule
         Construals & 933 & 41 \\
         \hspace{1em} Congruent (scene=fxn) & 803 & 25 \\ 
         \hspace{1em} Divergent (scene$\neq$fxn) & 130 & 16 \\ 
    \end{tabular}
    \caption{Statistics of the final Mandarin \emph{The Little Prince} Corpus (the Chinese SNACS Corpus). Tokenization, identification of adposition targets, and supersense labeling were performed manually.}
    \label{tab:stats}
\end{table}

Our corpus 
contains 933 manually identified adpositions. Of these, 70 distinct adpositions, 28 distinct scene roles, 26 distinct functions, and 41 distinct full construals are attested in annotation. 
Full statistics of token and type frequencies are shown in \Cref{tab:stats}.
This section presents the most frequent adpositions in Mandarin Chinese, as well as quantitative and qualitative comparisons of scene roles, functions, and construals between Chinese and English annotations.

\subsection{Adpositions in Chinese}\label{subsec: adpositiontokens}

We analyze semantic and distributional properties of adpositions in Mandarin Chinese.
The top 5 most frequent prepositions and postpositions are shown in \Cref{tab:stats_top_toks}.
Prepositions include canonical adpositions such as \textit{\yin1\wei4} and coverbs such as \textit{\zai4}. Postpositions are localizers such as \textit{\shang4} and \textit{\zhong1}.
We observe that prepositions \textit{\zai4} and \textit{\dui4} are dominant in the corpus (greater than 10\%). Other top adpositions are distributed quite evenly between prepositions and postpositions.
On the low end, 27 out of the 70 attested adposition types occur only once in the corpus.

\begin{table}[h!bt]
\centering\small
\begin{tabular}{llrrH}
\textbf{Prep.} & \textbf{Trans.} & \textbf{\%} & Count & \textbf{Type} \\
\hline
\zai4 & on & 18.4 & 172 & Prep \\
\dui4 & to & 11.0 & 103 & Prep \\
\ba3 & \textit{theme marker} & 7.2 & 67 & Prep \\
\yin1\wei4 & due to & 4.7 & 44 & Prep \\
\gei3 & to & 3.5 & 33 & Prep \\
\hline
\textbf{Total} & & 44.9 & 419 &  Prep\\

\hline\hline
\textbf{Postp.} & \textbf{Trans.} & \textbf{\%} & Count & \textbf{Type} \\
\shang4  & on top of & 9.5 & 89 & Post \\
\zhong1 & in the middle of & 4.9 & 46 & Post \\
\li3  & inside of & 3.9 & 36 & Post \\
\lai2\shuo1 & to one's regard & 2.7 & 25 & Post \\
\shi2 & at the time of & 2.1 & 20 & Post  \\
\hline
\textbf{Total} & & 23.2 & 216 & Post \\

\end{tabular}
\caption{Percentages and counts of the top 5 prepositions and postpositions in Chinese \textit{Little Prince}. The percentages are out of all adpositions.
}
\label{tab:stats_top_toks}
\end{table}


\subsection{Supersense \& Construal Distributions in Chinese versus English}

The distribution of scene role and function types in Chinese and English reflects the differences and similarities of adposition semantics in both languages. In \cref{tab:stats_supersense_zh_en} we compare this corpus with the largest English adposition supersense corpus, STREUSLE version~4.1 \citep{schneider-etal-2018-comprehensive}, which consists of web reviews.
We note that the Chinese corpus is proportionally smaller than the English one in terms of token and adposition counts.\footnote{We exclude possessives and multi-word expressions that are annotated in the English corpus since possessives are not formed by adpositional phrases in Mandarin Chinese.}
Moreover, there are fewer scene role, function and construal types attested in Chinese. The proportion of construals in which the scene role differs from the function (scene$\neq$fxn) is also halved in Chinese.
In this section, we delve into comparisons regarding scene roles, functions, and full construals between the two corpora both quantitatively and qualitatively. 


\begin{table*}[h!bt]
\centering\small
\begin{tabular}{lc|cc|cc|ccc}
 & \textbf{toks} &  \% \textbf{adps} & \textbf{uniq adps} &  \textbf{uniq scene} & \textbf{uniq fxn} &  \textbf{uniq cons} & \textbf{scene$\neq$fxn} & \textbf{\% scene$\neq$fxn}   \\
\midrule
\textbf{Chinese:} \textit{Little Prince} & 20k & 4.6\%  & \hphantom{1}70 &  28 & 26 & \hphantom{1}41 & \hphantom{0}16 & 14\% \\
\textbf{English:} \textit{EWT Reviews} & 55k &  7.4\% & 111 & 47 & 40 & 170 & 130  & 27\% \\
\end{tabular}
\caption{Statistics of Adpositional Supersenses in Chinese versus English. \textit{\% adps} presents the proportion of adposition targets over all token counts; \textit{uniq adps/scene/fxn/cons} demonstrates the type frequency of adposition tokens, scene role and function supersense and construals; \textit{scene$\neq$fxn} and \textit{\% scene$\neq$fxn} shows the type frequency and proportion of divergent construals.}
\label{tab:stats_supersense_zh_en}
\end{table*}

\paragraph{Overall Distribution of Supersenses}

\begin{figure}[h!bt]
    \centering
    \includegraphics[width=0.45\textwidth]{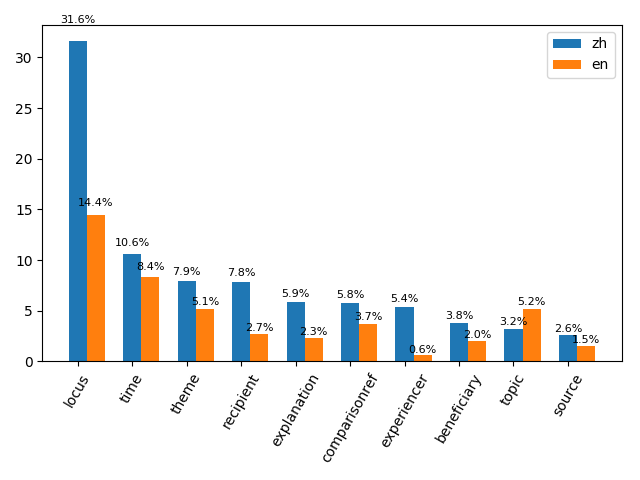}
    \caption{Top 10 most frequent scene roles in \textcolor{myblue}{Chinese} versus \textcolor{myorange}{English}.}
    \label{fig:bar_scene_zh_en}
\end{figure}

\begin{figure}[h!bt]
    \centering
    \includegraphics[width=0.45\textwidth]{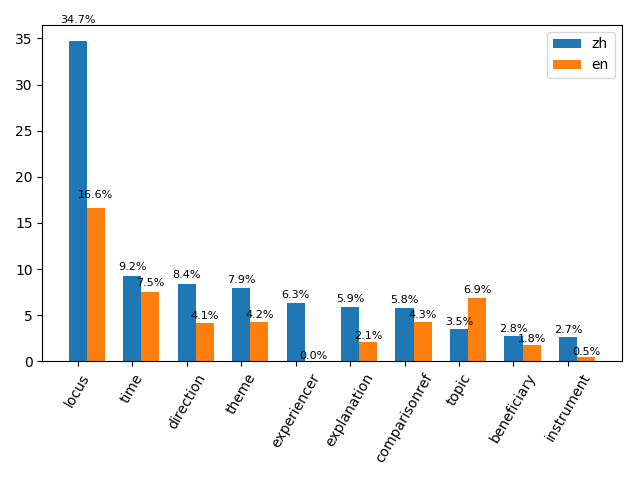}
    \caption{Top 10 most frequent functions in \textcolor{myblue}{Chinese} versus \textcolor{myorange}{English}.}
    \label{fig:bar_function_zh_en}
\end{figure}

\Cref{fig:bar_scene_zh_en,fig:bar_function_zh_en} present the top 10 scene roles and functions in Mandarin Chinese and their distributions in English. It is worth noting that since more scene role and function types are attested in the larger STREUSLE dataset, the percentages of these supersenses in English are in general lower than the ones in Chinese.  
\\[5pt]
There are a few observations in these distributions that are of particular interest. For some of the examples, we use an annotated subset of the English \textit{Little Prince} corpus for qualitative comparisons, whereas all quantitative results in English refer to the larger STREUSLE corpus of English Web Treebank reviews \citep{schneider-etal-2018-comprehensive}.

\paragraph{Fewer Adpositions in Chinese} 
As shown in 
\Cref{tab:stats_supersense_zh_en},
the percentage of adposition targets over tokens in Chinese is only half of that in English. This is due to the fact that Chinese has a stronger preference to convey semantic information via verbal or nominal forms. Examples~\cref{eg:en_moreadpositions,eg:zh_lessadpositions} show that the prepositions used in English, \textit{of} and \textit{in}, are translated as copula verbs (\textit{\shi4}) and progressives (\textit{\zheng4\zai4}) in Chinese. Corresponding to \Cref{fig:bar_scene_zh_en,fig:bar_function_zh_en}, the proportion of the supersense label \psst{Topic} in English is higher than that in Chinese; and similarly, the supersense label \psst{Identity} is not attested in Chinese for either scene role or function.

\ex. It was a picture \textbf{of:\psst{Topic}} a boa constrictor \textbf{in:\psst{Manner}} the act \textbf{of:\psst{Identity}} swallowing an animal . (\texttt{en\_lpp\_1943.3})
\label{eg:en_moreadpositions}

\exg. 
[\hua4 de] \textbf{\shi4} [[\yi4 \tiao2 \mang3\she2] \textbf{\zheng4\zai4} \tun1\shi2 [\yi4 \zhi1 \da4 \ye3\shou4]] \\
draw \textsc{de} \textsc{cop} one \textsc{cl} boa \textsc{prog} swallow one \textsc{cl} big animal \\
\trans `The drawing is a boa swallowing a big animal'. (\texttt{en\_lpp\_1943.3})
\label{eg:zh_lessadpositions}

\paragraph{Larger Proportion of \psst{Locus} in Chinese} In both \Cref{fig:bar_scene_zh_en} and \Cref{fig:bar_function_zh_en}, the percentages of \psst{Locus} as scene role and function are twice that of the English corpus respectively. This corresponds to the fact that fewer supersense types occur in Mandarin Chinese than in English. As a result, generic locative and temporal adpositions, as well as adpositions tied to thematic roles, have larger proportions in Chinese than in English.

\paragraph{\psst{Experiencer} as Function in Chinese}
Despite the fact that there are fewer supersense types attested in Chinese, \psst{Experiencer} as a function is specific to Chinese as it does not have any prototypical adpositions in English \citep{schneider2018guideline}. In \cref{eg:enexperiencergoal}, the scene role \psst{Experiencer} is expressed through the preposition \textit{to} and construed as \psst{Goal}, which highlights the abstract destination of the `air of truth'. This reflects the basic meaning of \textit{to}, which denotes a path towards a goal \citep{bowerman2001shaping}.
In contrast, the lexicalized combination of the preposition \textit{\dui4} and the localizer \textit{\lai2\shuo1} in \cref{eg:zhexperiencershenghuo} are a characteristic way to introduce the mental state of the experiencer, denoting the meaning `to someone's regard'. The high frequency of \textit{\lai2\shuo1} and the semantic role of \psst{Experiencer} (6.3\%)  underscore its status as a prototypical adposition usage in Chinese.

\ex. \textbf{To:\rf{Experiencer}{Goal}} those who understand life, that would have given a much greater air of truth to my story. (\texttt{en\_lpp\_1943.185})
\label{eg:enexperiencergoal}

\exg. 
[\textbf{\dui4:\psst{Experiencer}} [\dong3\de2 \sheng1\huo2 de \ren2] \textbf{\lai2\shuo1:\psst{Experiencer}}], \zhe4\yang4 \shuo1 \jiu4 \xian3\de2 \zhen1\shi2 \\
\textsc{p}:to know-about life \textsc{de} people \textsc{lc}:one's-regard this-way tell \textsc{res} seems real \\
\trans `It looks real to those who know about life.' (\texttt{zh\_lpp\_1943.185})
\label{eg:zhexperiencershenghuo}

\begin{figure}[h!bt]
    \centering
    \includegraphics[width=0.45\textwidth]{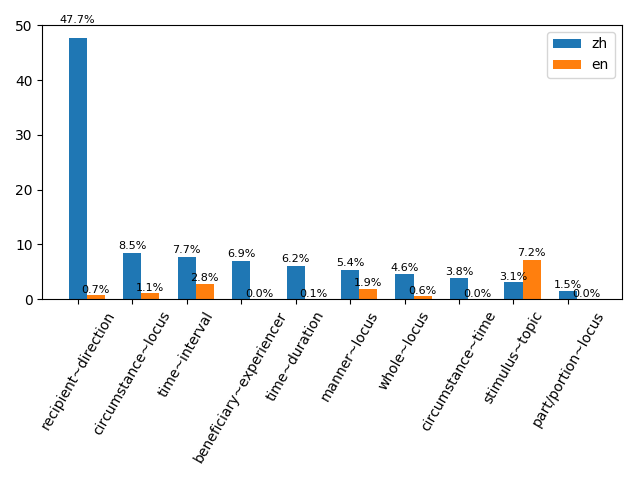}
    \caption{Top 10 Construals where scene$\neq$function in \textcolor{myblue}{Chinese} versus \textcolor{myorange}{English}.}
    \label{fig:bar_construal_zh_en}
\end{figure}

\paragraph{Divergence of Functions across Languages}
Among all possible types of construals between scene role and function, here we are only concerned with construals where the scene role differs from the function (scene$\neq$fxn). The basis of \citeposs{hwang-etal-2017-double} construal analysis is that a scene role is construed as a function to express the contexual meaning of the adposition that is different from its lexical one.
\Cref{fig:bar_construal_zh_en} presents the top 10 divergent (scene$\neq$fxn) construals in Chinese and their corresponding proportions in English. Strikingly fewer types of construals are formed in Chinese. Nevertheless, Chinese is replete with \rf{Recipient}{Direction} adpositions, which constitute nearly half of the construals. 
\\[5pt]
The 2~adpositions annotated with \rf{Recipient}{Direction} are \textit{\dui4} and \textit{\xiang4}, both meaning `towards' in Chinese. 
In \cref{eg:enrecipient,eg:zhrecipientdirection}, both English \textit{to} and Chinese \textit{\dui4} have \psst{Recipient} as the scene role. In \cref{eg:enrecipient}, \psst{Goal} is labelled as the function of \textit{to} because it indicates the completion of the ``saying'' event.\footnote{The prototypical function of \textit{to} indicates telic motion events. Telicity, however, is not required for \psst{Direction}.
} 
In Chinese, \textit{\dui4} has the function label \psst{Direction} provided that \textit{\dui4} highlights the orientation of the message uttered by the speaker as in \cref{eg:zhrecipientdirection}. Even though they express the same scene role in the parallel corpus, their lexical semantics still requires them to have different functions in English versus Chinese.

\ex. You would have to say \textbf{to:\rf{Recipient}{Goal}} them: ``I saw a house that costs \$$20,000$.'' (\texttt{en\_lpp\_1943.172}). \label{eg:enrecipient}

\exg. (\ni3) \bi4\xu1 [\textbf{\dui4:\rf{Recipient}{Direction}} \ta1men] \shuo1: ``\wo3 \kan4\jian4 le \yi2 \dong4 \shi2\wan4 \fa3\lang2 de \fang2zi.''\\
\textsc{2sg} must \textsc{P}:to \textsc{3pl} say \textsc{1sg} see \textsc{asp} one CL $10,000$ franc \textsc{de} house\\
`You must tell them: ``I see a house that costs 10,000 francs.'' ' (\texttt{zh\_lpp\_1943.172}). \label{eg:zhrecipientdirection}

\paragraph{New Construals in Chinese} 

Similar to the distinction between \rf{Recipient}{Goal} and \rf{Recipient}{Direction} in English versus Chinese, language-specific lexical semantics contribute to unique construals in Chinese, i.e.~semantic uses of adpositions that are unattested in the STREUSLE corpus. 
Six construals are newly attested in the Chinese corpus:

\begin{itemize}[noitemsep,topsep=0pt]
    \item \rf{Beneficiary}{Experiencer}
    \item \rf{Circumstance}{Time}
    \item \rf{PartPortion}{Locus}
    \item \rf{Topic}{Locus}
    \item \rf{Circumstance}{Accompanier}
    \item \rf{Duration}{Instrument}
\end{itemize}

Of these new construals, \rf{Beneficiary}{Experiencer} has the highest frequency in the corpus. The novelty of this construal lies in the possibility of \psst{Experiencer} as function in Chinese, shown by the parallel examples in \cref{eg:en_beni_beni,eg:zh_beni_expe}, where \textit{\dui4} receives the construal annotation \rf{Beneficiary}{Experiencer}. 

\ex. One must not hold it \textbf{against:\psst{Beneficiary}} them~. \texttt{(en\_lpp\_1943.180)}\label{eg:en_beni_beni}

\exg. 
\xiao3\hai3zimen \textbf{\dui4:\rf{Beneficiary}{Experiencer}} \da4\ren2men \ying4\gai1 \kuan1\hou4 xie \\
children P:to adults should lenient \textsc{comp} \\
\trans `Children should not hold it against adults.' \texttt{(zh\_lpp\_1943.180)}\label{eg:zh_beni_expe}

Similarly, other new construals in Chinese resulted from the lexical meaning of the adpositions that are not equivalent to those in English. For instance, the combination of \textit{\dang1 ... \shi2} (during the time of) denotes the circumstance of an event that is grounded by the time (\textit{\shi2}) of the event.  Different lexical semantics of adpositions necessarily creates new construals when adapting the same supersense scheme into a new language, inducing newly found associations between scene and function roles of these adpositions. Fortunately, though combinations of scene and function require innovation when adapting SNACS into Chinese, the 50 supersense labels are sufficient to account for the semantic diversity of adpositions in the corpus.

\section{POS Tagging of Adposition Targets}
\label{sec:adpositionidentification}

\begin{figure}[h!bt]
    \centering
    \includegraphics[width=0.35\textwidth]{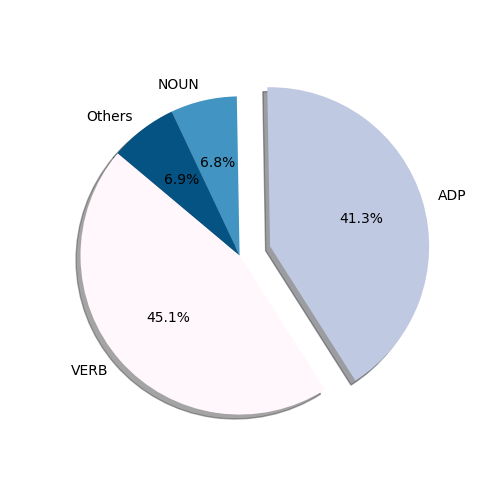}
    \caption{POS Distribution of Gold Adposition Tokens.}
    \label{fig:pie_gold_pos}
\end{figure}

We conduct a post-annotation comparison between manually identified adposition targets and automatically POS-tagged adpositions in the Chinese SNACS corpus. Among the 933 manually identified adposition targets that merit supersense annotation, only 385 (41.3\%) are tagged as \pos{adp} (adposition) by StanfordNLP \citep{qi2018universal}.
\Cref{fig:pie_gold_pos} shows that gold targets are more frequently tagged as \pos{verb} than \pos{adp} in automatic parses, as well as a small portion that are tagged as \pos{noun}. The inclusion of targets with \pos{pos=verb} reflects our discussion in \cref{subsec:adposition_criteria} that coverbs co-occurring with a main predicate are included in our annotation. 
The automatic POS tagger also wrongly predicts some non-coverb adpositions, such as \textit{\yin1\wei2}, to be verbs.

The StanfordNLP POS tagger also suffers from low precision (72.6\%). Most false positives resulted from the discrepancies in adposition criteria between theoretical studies on Chinese adpositions and the tagset used in Universal Dependencies (UD) corpora such as the Chinese-GSD corpus. For instance, the Chinese-GSD corpus considers subordinating conjunctions (such as \textit{\ru2\guo3}, \textit{\yi2\dan4}, \textit{\ji4\ran2}, \textit{\zhi3\yao4}) adpositions; however, theoretical research on Chinese adpositions such as \citet{li1989mandarin} differentiates them from adpositions, since they can never syntactically precede a noun phrase. 
\\[5pt]
Hence, further SNACS annotation and disambiguation efforts on Chinese adpositions cannot rely on the StanfordNLP \pos{adp} category to identify annotation targets. 
Since adpositions mostly belong to a closed set of tokens, we apply a simple rule to identify all attested adpositions which are not functioning as the main predicate of a sentence, i.e., not the \textit{root} of the dependency tree. 
As shown in Table \ref{tab:heuristic}, our heuristic results in an $F_1$ of 82.4\%, outperforming the strategy of using the StanfordNLP POS tagger. 

\begin{table}[h!bt]
\centering
\begin{tabular}{lccc}
& $P$ & $R$ & $F_1$ \\
\midrule
StanfordNLP \textsc{ADP} & 72.6  & 41.3 & 52.6 \\
attested \texttt{dep}$\neq$\texttt{root} adpositions & 75.1 & 91.3 & 82.4 \\
\end{tabular}
\caption{Adposition identification performance on Chinese SNACS corpus.}
\label{tab:heuristic}
\end{table}

\section{Conclusion}

In this paper, we presented the first corpus annotated with adposition supersenses in Mandarin Chinese. 
The corpus is a valuable resource for examining similarities and differences between adpositions in different languages with parallel corpora and can further support automatic disambiguation of adpositions in Chinese. 
We intend to annotate additional genres---including native (non-translated) Chinese and learner corpora---in order to more fully capture the semantic behavior of adpositions in Chinese as compared to other languages.

\finalversion{\section*{Acknowledgements}
We thank 
anonymous reviewers for their feedback.
This research was supported in part by NSF award IIS-1812778 and grant 2016375 from the United States--Israel Binational Science Foundation (BSF), Jerusalem, Israel.
}


\bibliographystyle{lrecnat}
\bibliography{snacs}


\newpage









\end{document}